\documentclass[apj]{emulateapj}
\pdfoutput=1
\usepackage{amsmath}
\usepackage{mathrsfs}

\newcommand{\msun}{$M_{\odot}$}

\newcommand{\eg}{{\it e.g.}}
\newcommand{\ie}{{\it i.e.}}
\newcommand{\cmt}{cm$^{-3}$}

\begin{document}
\slugcomment{accepted for publication into PASP}
\shorttitle{Discerning the Form of the Dense Core Mass Function}
\shortauthors{Swift \& Beaumont}
\title{Discerning The Form of the Dense Core Mass Function}
\author{Jonathan J. Swift \& Christopher N. Beaumont}
\email{js@ifa.hawaii.edu, beaumont@ifa.hawaii.edu}
\affil{Institute for Astronomy, 2680 Woodlawn Dr., Honolulu, HI
  96822-1897}

\begin{abstract}
We investigate the ability to discern between lognormal and powerlaw
forms for the observed mass function of dense cores in star forming
regions. After testing our fitting, goodness-of-fit, and model
selection procedures on simulated data, we apply our analysis to
14 datasets from the literature. Whether the core mass function
has a powerlaw tail or whether it follows a pure lognormal form cannot
be distinguished from current data. From our simulations it is
estimated that datasets from uniform surveys containing more than 
$\approx 500$ cores with a completeness limit below the peak of the
mass distribution are needed to definitively discern between these two 
functional forms. We also conclude that the width of the core mass
function may be more reliably estimated than the powerlaw index of the
high mass tail and that the width may also be a more useful parameter in
comparing with the stellar initial mass function to deduce the
statistical evolution of dense cores into stars.
\end{abstract}

\keywords{stars: formation --- ISM: clouds ---  ISM: structure}

\section{Background} \label{Sintro}
Stars form from dense cores of molecular gas and dust
with sizes $\approx 0.1$\,pc, masses between $\approx 0.5$ and
50\,\msun, and densities $\gtrsim 10^4$\,\cmt\
\citep[see, {\eg},][]{ber07}. Therefore the relationship between mass
distributions of dense cores in star forming regions (core mass
function, or CMF) and the stellar initial mass function (IMF) 
contains information regarding how observed samples of cores evolve
into stars. The observational similarity between the CMF and the IMF
was first put forth by \cite{mot98}, and since this time many other
samples of dense cores have been presented in this context
\citep[{\eg},][]{sim08,eno08,alv07,nut07,stan06,joh01}. The
qualitative similarity between the CMF and the IMF offers support for
the accepted idea that stars form from dense cores
\citep[{\eg}][]{lad08}. But what can we learn about the star formation
process from this observed similarity?

In a recent paper, we presented a series of numerical experiments
evolving distributions of cores into stellar IMFs
\citep{swi08b}. We find that a given CMF
evolved according to different evolutionary pathways produces
variations in the resultant IMF that are insignificant in
relation to the errors inherent in current samples of dense cores. Our
results show that the form of the CMF in relation to the IMF indeed
contains vital clues to the star formation process, but highlight the
difficulty in deducing how observed samples of cores evolve into stars
based solely on the shapes of these distributions.

The central limit theorem applied to isothermal turbulence naturally
predicts a lognormal probability distribution in density
\citep{lar73,zin84,ada96} that has been produced in computer
simulations \citep[{\eg},][]{vaz94,nor99,kle01}. Extensive surveys of
nearby star-forming regions have also shown a CMF shape consistent
with a lognormal form \citep[{\eg},][]{eno08,stan06}.

However, observed CMFs are typically characterized by one or more
powerlaws. While the use of a powerlaw form to fit observed CMFs has
never been rigorously justified, this form is assumed based on its
versatility and the expected similarity between the CMF and the IMF 
\citep[{\eg},][]{mot98}. Recent data show that this apparent powerlaw
behavior does not extend to very low masses but displays a turnover or 
break below a few \msun\ \citep{alv07}. Motivated by
the observational similarity of the CMF and the IMF, theories have
been recently developed describing physical scenarios that may produce
a powerlaw distribution of molecular core masses that turns over
toward low-masses \citep{pad02,shu04b,hen08}.

To understand how, or if, dense molecular cores produce the full
spectrum of stellar masses, it is essential to understand the
probability distribution function (PDF) from which the CMF is
drawn. If the parent distribution has a powerlaw tail extending to
high masses, this would support the idea that stellar mass is almost
entirely determined in the molecular cloud phase
\citep[{\eg},][]{alv07}. If observed mass distributions of cores are
found to be drawn from a purely lognormal
PDF, the origin of the powerlaw distribution of stellar masses will
remain an open question. A lognormal CMF would disfavor the idea that
massive stars form directly from massive cores 
\citep[such as][]{kru09}, and may imply that massive stars form
through mechanisms distinct from low-mass stars
\citep[{\eg},][]{bon04}. Distinguishing between these two forms is
complicated by the difficulty in measuring the CMF over large dynamic
ranges and the fact that lognormal and powerlaw forms can look quite
similar over limited mass ranges.

In 2006, Reid \& Wilson concluded that observed CMFs are consistent
with being drawn from the same parent distribution and that this
parent distribution is consistent with the IMF. However, the results
of their Kolmogorov-Smirnov (KS) tests applied to the median
normalized CMFs suggest that the CMFs are indistinguishable while
their goodness-of-fit statistic based on $\chi^2$ suggest that
different core samples may have different preferred forms. Given these
ambiguities and the fact that the KS test is somewhat
compromised when applied to two distributions with normalized medians,
we revisit the topic of discerning the form of the parent
distribution to the CMF with a new approach. A renewed interest in
this subject is motivated by new instruments coming online ({\eg},
Herschel, SCUBA2, ALMA) and surveys that will provide large, uniform
samples of carefully selected cores\footnote{{\eg}, The Gould's Belt
Legacy Survey\\ {\tt http://www.jach.hawaii.edu/JCMT/surveys/gb}}.

In \S\,\ref{Sdata} we outline our analysis strategy and then test our
procedures on simulated data in \S\,\ref{Ssim}. The application of our
procedures to 14 state-of-the-art datasets of dense cores in
star-forming regions is presented in
\S\,\ref{Sreal}, and we close with a discussion of these results in
\S\,\ref{Sdiscussion} that includes suggestions for future
observations. 

\section{Data Analysis} \label{Sdata}
The goals of our data analysis are to (1) find the best fit parameters for
lognormal and powerlaw models given the data, (2) assess whether or
not a given model adequately describes the data, (3) select either the
lognormal or powerlaw model as the preferred model, and (4) compute
the confidence in that selection. The following sub-sections outline
the details of each step used to achieve these goals. All procedures
were written in the Interactive Data Language\footnote{IDL;
{\tt http://www.ittvis.com/ProductServices/IDL.aspx}}.   

\subsection{The Models} \label{SPDF}
The two models we choose to differentiate in our analysis are
motivated in \S\,\ref{Sintro} and are described by probability
distribution functions (PDFs) of powerlaw and lognormal form. 
The powerlaw PDF is given by
\begin{equation}
p_{\rm pl}(m) =
C_{\rm pl}m^{-\alpha} 
\label{pl}
\end{equation}
for $m \ge m_0$. The normalization constant is given by
\begin{equation*}
C_{\rm pl} = (\alpha - 1)m_0^{\alpha -1}
\label{Cpl}
\end{equation*}
where $m_0$ signifies the break or ``turnover'' below which the
distribution does not follow powerlaw behavior. 

The lognormal PDF is given by
\begin{equation}
p_{\rm ln}(m) = \frac{C_{\rm ln}}{m}\exp\left[-\frac{(\ln m -
    \mu)^2.}{2\sigma^2}\right]
\label{ln}
\end{equation}
with the normalization constant
\begin{equation*}
C_{\rm ln} = \sqrt{\frac{2}{\pi\sigma^2}}\left[{\rm erfc}\left(\frac{\ln
    m_{\rm lim} - \mu}{\sigma\sqrt{2}}\right)\right]
\label{Cln}
\end{equation*}
where the characteristic mass $m_{\rm c} = e^\mu$ and  $m_{\rm lim}$
is the minimum mass of the distribution, analogous to an
observational completeness limit.  

\subsection{Fitting} \label{Sfit}
To avoid the dangers inherent in fitting data using regression models
arising from, \eg, data binning \citep{ros05}, we use the method of
maximum likelihood to estimate the model parameters for both simulated and
real data \citep[see][]{claus07}.

The best estimate of the powerlaw index for data with values above
$m_0$ is found analytically by maximizing the log-likelihood:
\begin{equation}
\hat{\alpha} = 1+ n\left[\sum_{i = 1}^{n}\ln\left(\frac{m_i}{m_0}\right)\right]^{-1}
\label{alpha}
\end{equation}
where $n$ is the number of data with $m \ge m_0$. Hatted quantities
will signify the best fit parameters throughout the text. The value of
$m_0$ is also unknown and is found by minimizing the KS statistic
between the best fit model and the data as a function of $m_0$
\citep[][\S\,3.3]{claus07}. 

The likelihood for the distribution in Equation \ref{ln} cannot be
maximized analytically. Therefore, we numerically maximize the
likelihood for the lognormal model as a function of $\sigma$ and $\mu$
using Powell's method \citep{pre07}.

\subsection{Goodness-of-fit} \label{Sgood}
Once the data are fit, we calculate the KS statistic between the best
fit model and the data as a proxy for goodness-of-fit. When one or
more parameters of a model are 
determined from data, the probability value for the KS statistic,
$p_{\rm KS}$, can no longer be found directly
from the KS probability distribution \citep{lil67}. We therefore
calculate $p_{\rm KS}$ by relating the KS statistic to the
distribution of values generated from 5000 parametric bootstrap 
realizations \citep{pre07,bab06,claus07}. 

The value of $p_{\rm KS}$ represents the probability of getting a fit
{\em worse} than the original given the model and the PDF of the
distribution. Low values of $p_{\rm KS}$, typically below 0.1 or 0.05,
are a flag for a given model being an inadequate description of the
data.

\subsection{Model Selection} \label{Smodel}
There are several approaches to selecting between models. Bayesian
model selection compares the evidence of each model computed as an
integral over all free parameters of the probability density, {\ie},
the likelihood function. One of our free parameters, $m_0$, determines
the number of data points to be considered by the models thus adding a 
difficult complexity to this approach. Instead we use the likelihood
ratio test, a similar approach that compares the peak probability
densities of the models. Taking the log of this ratio we obtain the
parameter 
\begin{equation}
\mathcal{R} = \sum_{i = 1}^{n}\left[\ln p_{\rm pl}(m_i) - \ln p_{\rm ln}(m_i)\right]
\label{lratio}
\end{equation}
where positive values of $\mathcal{R}$ favor the powerlaw model and
negative values favor the lognormal model. 

Our two models each have two parameters---$m_0$ and $\alpha$ for the
powerlaw model, and $\mu$ and $\sigma$ for the lognormal
model. Therefore we do not expect there to be biases introduced into
this method due to different levels of 
complexity in the models. An additional benefit to using this test is
that the significance of ${\rm sgn}(\mathcal{R})$ can be assessed by
generating a distribution of $\mathcal{R}$ values through bootstrap
resampling of the data (for real data) or the PDF (for simulated
data). The fraction of $\mathcal{R}$ values having the same sign as
the original value is designated $C_{\mathcal{R}}$, the confidence
level of the likelihood ratio.  

\section{Application to Simulated Data}\label{Ssim}
To ascertain how well our procedures discriminate between powerlaw and
lognormal distributions, we have carried out our analysis on a large
number of simulated datasets. These simulations quantify how sensitively
the model selection process depends on the sample size and
$m_{\rm lim}$. 

Distributions with $N_{\rm C}$ samples above a completeness limit of
$m_{\rm lim}$ for $N_{\rm C} = 50\times2^n$ ($n = 0,1,2,...,8$) and
$m_{\rm lim} =$ 0.1, 0.5, and 1.0\,\msun\ are drawn from the PDFs of
\S\,\ref{SPDF}. For powerlaw data, we draw from the distribution in
Equation \ref{pl} for masses above $m_0$ while for masses between
$m_{\rm lim}$ and $m_0$ we draw from Equation \ref{ln}. The powerlaw
data are drawn such that the parent distribution is smooth and
continuous at $m_0 = \mu - \sigma^2(1-\alpha)$. Lognormal data are
drawn from the distribution in Equation \ref{ln}. For each sample size,
$N_{\rm c}$, and completeness limit, $m_{\rm lim}$, 5000 distributions
are realized and fit according to the procedures outlined in
\S\,\ref{Sdata}.  

We use values $\sigma = 1.0$ and $\mu = 0$ for all simulated data,
consistent with CMFs measured in nearby star-forming regions
\citep{eno08,stan06}. For powerlaw data we use $\alpha = 2.5$, close
to the Salpeter value of 2.35 \citep{sal55} and also consistent with
data from the literature 
\citep[][also see Table~\ref{fittable}]{rei06b}. 

\subsection{Likelihood Ratio Test} \label{LR}
\begin{figure}[!t]
\centering
\includegraphics[angle=0,width=3.4in]{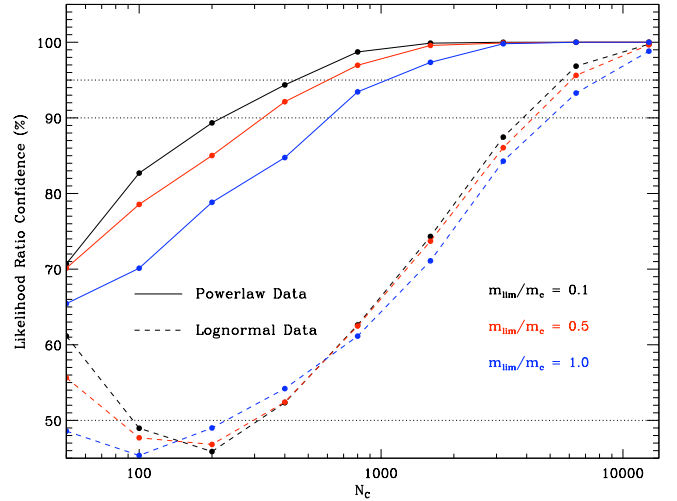}
\caption{The confidence with which the likelihood ratio test can
  discern between a powerlaw and lognormal model for data generated
  from a powerlaw ({\em solid lines}) and lognormal ({\em dashed
  lines}) probability distribution function as a function of the
  sample size, $N_{\rm C}$. Data with completeness
  limits of 0.1, 0.5, and 1.0 times the characteristic mass are shown
  in {\em black}, {\rm red}, and {\em blue}, respectively.}
\label{R_Conf}
\end{figure}

Figure~\ref{R_Conf} displays $C_{\mathcal{R}}$ as a function of
$N_{\rm C}$ given different values of $m_{\rm lim}$. The solid lines
in the figure represent the confidence with which the underlying
powerlaw form to the PDF can be discerned from a lognormal form. The
confidence is an increasing function of $N_{\rm C}$ and a 
decreasing function of 
$m_{\rm lim}$ expressed here as the fraction of the characteristic
mass of the distribution, $m_{\rm lim}/m_{\rm c}$. The dependency on
$N_{\rm C}$ is expected and reflects the increasing amount of
information available to the likelihood ratio test. The dependency on
$m_{\rm lim}$ arises because a lognormal form can closely
approach a powerlaw form over a 
limited range for large negative values of $\hat{\mu}$ and large
positive values of $\hat{\sigma}$. Therefore, as more data are
included below $m_0$ stronger constraints can be placed on the values
of $\hat{\mu}$ and
$\hat{\sigma}$ thereby creating a larger discrepancy between a
lognormal model and the high mass tail of the 
simulated data. We find that a lognormal form to the CMF can be ruled
out at the 95\% confidence level for a survey of intermediate
sensitivity and sample sizes greater than $\approx 500$. 

The dashed lines in Figure~\ref{R_Conf} represent the confidence level
with which a pure lognormal form to the PDF can be distinguished from
a one with a powerlaw tail. We see that these curves are monotonically
increasing for $N_{\rm C}\gtrsim 100$. The upturn in the lognormal
data curves at low  $N_{\rm C}$ is due to the effects of fitting a
powerlaw to a small number of data points. As $N_{\rm C}$ increases,
so do the best fit values, $\hat{m_0}$ and $\hat{\alpha}$, thereby
making a distinction between models more difficult than for the case
of powerlaw data. If $m_0$ is forced to lie at some fraction of
$m_{\rm c}$, then results similar to the a powerlaw data 
({\em solid lines}) are obtained. The dependency of $C_{\mathcal{R}}$
on $m_{\rm lim}$ is weaker for the lognormal data than the powerlaw
data since lower values of $m_{\rm lim}$ produce only slightly better
lognormal fits. 

\subsection{Goodness-of-fit Test}
\begin{figure}[!b]
\centering
\includegraphics[angle=0,width=3.4in]{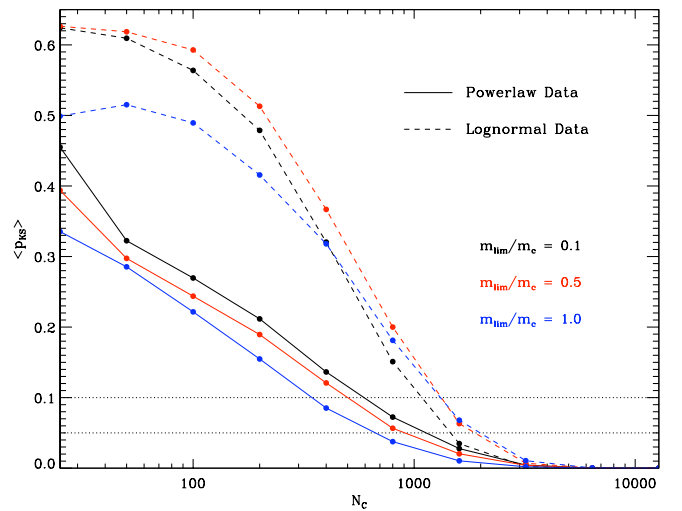}
\caption{The average KS probability resulting from fitting a lognormal
  model to powerlaw data ({\em solid lines}) or a powerlaw model to
  lognormal data ({\em dashed lines}). Color conventions are the same
  as in Figure~\ref{R_Conf}.} 
\label{pKS}
\end{figure}

Figure \ref{pKS} shows the average value of $p_{\rm KS}$ as a function
of $N_{\rm C}$ generated by fitting the incorrect model to the
simulated data. For both the powerlaw and lognormal data, the
incorrect model produces poorer fits as the sample size
increases. Similar to the likelihood ratio test, it is more difficult
to rule out a powerlaw model from lognormal data. However, the
discrepancy between the critical values of $N_{\rm C}$ is smaller
using this test.

For powerlaw data, $\langle p_{\rm KS}\rangle$ is a decreasing
function of $m_{\rm lim}$. This dependency comes about since the KS
test is most sensitive to the median of a distribution and the
powerlaw and lognormal models have median values that become closer as
$m_{\rm lim}$ decreases. For lognormal data,
$\langle p_{\rm KS}\rangle$ is not a monotonic function of $m_{\rm lim}$ for
all values of $N_{\rm C}$.

It may be expected that the curves of $\langle p_{\rm KS}\rangle$
should approach 0.5 as $N_{\rm C} \rightarrow 0$, meaning that the two
models are indistinguishable for small sample sizes. However, it can
be seen that the curves representing lognormal data ({\em dashed
lines}) overshoot this value for $N_{\rm C}$ between about 50 and
200. As in \S\,\ref{LR}, this results from fitting a powerlaw to a
small number of samples and represents a bias for lognormal data to
look marginally more like a powerlaw for $N_{\rm C}$ in this range
according to our goodness-of-fit test. 

From these simulations it can be concluded that fitting a powerlaw
model to lognormal data or conversely a lognormal model to powerlaw
data will consistently yield very low values of $p_{\rm KS}$ for
$N_{\rm C}$ greater than several hundred to a thousand. 

\section{Application to Literature Data} \label{Sreal}
Table~\ref{littable} lists the 14 datasets to which we apply the
analysis of \S\,\ref{Sdata} along with their descriptive
parameters. Various techniques are used to observationally identify
the cores in these datasets. The most common technique is the use
of threshold contouring \citep[{\eg},][]{alv07}, though gaussian
fitting \citep[{\eg},][]{nut07} and wavelet decomposition
\citep[{\eg},][]{mot98} are also used.

The results of our fitting procedures, goodness-of-fit
tests, and model selection techniques are displayed in
Table~\ref{fittable}. For a majority of the datasets, neither a
lognormal form nor a powerlaw provide a significantly better
description of the data. Five datasets out of 14 have values of
$p_{\rm KS}$ and $C_{\mathcal R}$ that are potentially
significant. These values are underlined in the table for emphasis. 

Datasets 3, 11 and 14 disfavor a powerlaw model due to a very
low $p_{\rm KS}$ value while showing various levels of support for a
lognormal model in the likelihood ratio. Dataset 12 disfavors a
lognormal form in the $p_{\rm KS}$ statistic while showing moderate
support for a powerlaw form in the likelihood ratio test. Dataset 6
is not adequately fit by either model. 
\begin{center}
\begin{deluxetable}{lccccc}[!hb]
\tablewidth{0in}
\tablecaption{Literature Data}
\tablehead{
\colhead{Dataset ID} &
\colhead{Region} &
\colhead{$N_{\rm tot}$} &
\colhead{$m_{\rm lim}$} &
\colhead{$N_C$} &
\colhead{Ref.}}
\startdata
1\dotfill & Pipe Nebula         & 159 & 1.0 &  73 & 1 \\ 
2\dotfill & Pipe Nebula         & 134 & 0.6 &  81 & 2 \\ 
3\dotfill & Bolocam Cores (SL)  & 108 & 0.8 &  69 & 3 \\
4\dotfill & Bolocam Cores (PS)  & 92  & 0.8 &  45 & 3 \\
5\dotfill & Bolocam Cores (All) & 200 & 0.8 & 114 & 3 \\
6\dotfill & $\rho$\,Ophiuchus   & 55  & 0.1 &  40 & 4 \\
7\dotfill & $\rho$\,Ophiuchus   & 143 & 0.1 & 120 & 5 \\
8\dotfill & $\rho$\,Ophiuchus   & 62  & 0.1 &  51 & 6 \\
9\dotfill & Orion A             & 172 & 0.3 & 170 & 7 \\
10\dotfill & Orion              & 203 & 1.0 & 149 & 8,9,10 \\
11\dotfill & M17                & 121 & 3.0 & 108 & 11 \\
12\dotfill & NGC\,7538          &  77 &  15 &  45 & 12 \\
13\dotfill & Cygnus X           & 129 & 5.3 & 125 & 13 \\
14\dotfill & NGC\,6334          & 181 &  32 & 131 & 14 \\
\enddata 
\tablerefs{(1) \cite{alv07}, (2) \cite{rat09}, (3) \cite{eno08}, (4)
  \cite{joh00}, (5) \cite{stan06}, (6) \cite{mot98}, (7)
  \cite{nut07}, (8) \cite{joh01}, (9)\cite{joh06a}, (10) \cite{joh06b}, (11) \cite{rei06a},
  (12) \cite{rei05}, (13) \cite{mot07}, (14) \cite{mun07}}
\label{littable}
\end{deluxetable}
\end{center}

\begin{center}
\begin{deluxetable*}{lccccccccccc}
\tablewidth{0in}
\tablecaption{Literature Analysis Results}
\tablehead{
\colhead{} &
\multicolumn{3}{c}{Powerlaw Fit} &
\colhead{} &
\multicolumn{3}{c}{Lognormal Fit} \\
\cline{2-4} \cline{6-8}
\colhead{Dataset ID} &
\colhead{$\hat{\alpha}$} &
\colhead{$\hat{m_0}$} &
\colhead{$p_{KS}$} &
\colhead{$N_{\rm tail}$\tablenotemark{a}} &
\colhead{$\hat{\mu}$}&
\colhead{$\hat{\sigma}$}& 
\colhead{$p_{KS}$} &
\colhead{} &
\colhead{$m_{\rm lim}/m_{\rm c}$} &
\colhead{$\mathcal{R}$} &
\colhead{$C_{\mathcal{R}}$}}
\startdata
1\dotfill  & $2.9^{+0.6}_{-0.5}$ & $2.5^{+0.04}_{-1.1}$ & 0.75 & 27 & $-0.1^{+0.6}_{-2.6}$ & $1.1^{+0.8}_{-0.3}$ & 0.90 & & 1.1 & PL & 76\% \\ 
2\dotfill  & $3.0^{+0.6}_{-0.5}$ & $2.4^{+0.3}_{-1.3}$  & 0.43 & 30 & $ 0.2^{+0.3}_{-0.6}$ & $1.0^{+0.3}_{-0.2}$ & 0.51 & & 0.5 & PL & 71\% \\ 
3\dotfill  & $2.7^{+0.3}_{-0.3}$ & $0.9^{+0.7}_{-0.1}$  & \underline{0.00} & 65 & $0.0^{+0.3}_{-0.8}$  & $0.7^{+0.3}_{-0.2}$ & 0.42 & & 0.8 & LN & \underline{91\%} \\
4\dotfill  & $3.4^{+0.5}_{-0.5}$ & $2.5^{+0.1}_{-0.9}$  & 0.72 & 22 & $0.7^{+0.2}_{-0.4}$  & $0.7^{+0.3}_{-0.1}$ & 0.68 & & 0.4 & PL & 68\% \\
5\dotfill  & $4.1^{+1.0}_{-0.8}$ & $2.7^{+0.1}_{-1.5}$  & 0.32 & 28 & $0.2^{+0.2}_{-0.5}$  & $0.8^{+0.2}_{-0.1}$ & 0.54 & & 0.7 & PL & 60\% \\
6\dotfill  & $1.9^{+0.2}_{-0.2}$ & $0.2^{+0.5}_{-0.01}$ & \underline{0.00} & 37 & $-0.9^{+0.2}_{-0.3}$ & $0.9^{+0.3}_{-0.2}$ & \underline{0.06} & & 0.2 & LN & 77\% \\
7\dotfill  & $2.8^{+0.5}_{-0.4}$ & $0.6^{+0.04}_{-0.2}$ & 0.85 & 43 & $-1.1^{+0.2}_{-0.3}$ & $1.0^{+0.2}_{-0.1}$ & 0.45 & & 0.3 & PL & 85\% \\
8\dotfill  & $2.0^{+0.2}_{-0.2}$ & $0.1^{+0.1}_{-0.02}$ & 0.15 & 46 & $-4.1^{+2.1}_{-2.0}$ & $1.9^{+0.8}_{-0.7}$ & 0.20 & & 5.9 & LN & 69\% \\
9\dotfill  & $1.9^{+0.2}_{-0.2}$ & $6.4^{+1.8}_{-5.3}$  & 0.67 & 37 & $-5.4^{+1.4}_{-3.9}$  & $3.4^{+1.1}_{-0.5}$ & 0.94 & & 200 & PL & 56\% \\
10\dotfill & $2.7^{+0.3}_{-0.3}$ & $3.1^{+2.4}_{-0.6}$  & 0.21 & 69 & $0.7^{+0.2}_{-0.4}$  & $1.0^{+0.2}_{-0.1}$ & 0.45 & & 0.5 & PL & 68\% \\
11\dotfill & $2.1^{+0.2}_{-0.2}$ & $14^{+6}_{-8}$      & \underline{0.00} & 59 & $2.6^{+0.2}_{-0.3}$  & $1.1^{+0.2}_{-0.2}$ & 0.59 & & 0.2 & LN & \underline{90\%} \\
12\dotfill & $2.1^{+0.4}_{-0.32}$ & $68^{+4}_{-45}$    & 0.57 & 22 & $2.7^{+1.1}_{-3.0}$  & $1.9^{+1.3}_{-0.6}$ & \underline{0.07} & & 1.0 & PL & 84\% \\
13\dotfill & $2.1^{+0.2}_{-0.2}$ & $21^{+5}_{-5}$      & 0.66 & 65 & $2.7^{+0.3}_{-0.5}$  & $1.3^{+0.3}_{-0.2}$ & 0.14 & & 2.2 & PL & 83\% \\
14\dotfill & $1.7^{+0.1}_{-0.1}$ & $47^{+41}_{-13}$    & \underline{0.09} & 111 & $2.6^{+1.3}_{-3.9}$  & $2.2^{+1.2}_{-0.5}$ & 0.74 & & 2.3 & LN & 74\% \\
\enddata 
\tablenotetext{a}{Number of cores with $m \ge m_0$.}
\label{fittable}
\end{deluxetable*}
\end{center}

\section{Discussion}\label{Sdiscussion}
\subsection{Does the CMF Have a Powerlaw Tail?}\label{Stail}
Ambiguous results for more than 60\% of the datasets and conflicting
results for the remaining $\sim 40$\% emphasize the difficulty in
discerning the form of the CMF. Given the results in
Table~\ref{fittable}, we conclude that a pure lognormal PDF and a PDF
with a powerlaw tail cannot be distinguished from current
core datasets. Furthermore, neither the existence
of a powerlaw tail to the CMF nor the uniformity of the CMF from
region to region are evident.

It is not clear to what degree systematic differences in each
dataset---differing observational techniques, physical resolution,
filtering, core identification algorithms, and parameters used in
deriving core mass ({\eg},  
$\kappa_\nu$)---contribute to the variations seen from dataset to
dataset, and it is possible that a uniform re-analysis of the datasets
could reveal a more definitive relationship between the shapes of the
CMFs. Also, the free parameters are unconstrained in our fitting
procedures, and in some cases this leads to fits that could be seen as 
unreasonable ({\eg}, the low values of $\hat{\mu}$ 
for datasets 8 and 9). Judicious constraints on the fitting parameters
could therefore lead to more decisive results. 

However, our conclusions are consistent with the results from 
\S\,\ref{Ssim}. The literature datasets of Table~\ref{littable} have
$N_{\rm C} \lesssim 150$ while the results of \S\,\ref{Ssim} suggest
that more than $\approx 500$ samples are needed to definitively
discern between these two models. These results highlight the need for
larger, uniform survey data to produce a statistically significant
result regarding the functional form of the CMF.

\subsection{Comparing the CMF to the IMF} \label{Scomp}
The utility in analyzing the form of the CMF is in its relation to the
stellar IMF to address the question of how the final mass of a star is
determined. The similarity of the CMF to the IMF has inspired the idea
that the masses of stars are determined via the fragmentation of dense
molecular gas in the 
Galaxy. It seems as though this must be true at least in part since it
is known that stars form in dense, self-shielded molecular gas. However,
whether or not the mass distributions of molecular cores observed in
star-forming regions will accurately represent the mass distribution
of stars to form from them is still debatable.

It was shown by \cite{swi08b} that different evolutionary pathways
from cores to stars produce variations in the form of the resultant
IMF. The powerlaw slope was shown to be quite robust except under the
most extreme evolutionary scenarios due to the scale-free nature of
powerlaw distributions. The width however was shown to be
a more sensitive indicator of core evolution. Given the small dynamic
range over which the CMF can be measured in nearby star-forming
regions, we emphasize the importance of obtaining a sound measurement
of $\sigma$ in survey data as
this parameter may be more powerful than 
$\alpha$ in constraining how observed distributions of cores evolve
into stars. 

The width of the IMF has been measured to be between 0.3 and 0.7\,dex
\citep{mil79,cha03,boc09}. The measured
values of $\hat{\sigma}$ in Table~\ref{fittable} tend to be higher
than this, approximately between 0.7 and 2. If this were confirmed to
be the case in a more sensitive, uniform survey, this would indicate
that an additional mass preference occurs in later stages of
gravitational collapse likely through fragmentation into several stars
at or near the local Jeans mass. 

\subsection{Bettering Our Understanding of the CMF}
It is a difficult task to automate the identification of large numbers
of pre-stellar molecular cores in star-forming regions, and there is
no consensus on how to define a core observationally. However,
methods are currently being improved \citep{kai09,pin09} and new
strategies are being developed \citep[{\eg},][]{ros08}. It has been
recently shown that the physical characteristics of cores
\citep[{\eg},][]{rat09}, the potential variability of core lifetimes
\citep{cla07} as well as the effects of statistical errors
\citep[{\eg},][]{koe09,ros05} are also important to consider when
interpreting observed CMFs.

From this work, we present three new
suggestions aimed to help maximize the utility of future surveys
of dense cores in star-forming regions. One, is that sample sizes of
greater than $\approx 500$ above the completeness limit should be
sought so that a lognormal and powerlaw tail to the CMF can be
discerned. The second suggestion is that the survey should be as
uniform as possible such that systematic errors can be minimized. The
third suggestion is that a sensitivity limit of better than one half
of the peak mass of the distribution is desirable such that the width
of the CMF can be reliably determined and compared to the width of the
IMF.

A survey such as the Gould's Belt Legacy Survey to be conducted with
SCUBA2 on the James Clerk Maxwell Telescope may very well reach these
goals, and the value of $\sigma$ is likely to be measured with
unprecedented accuracy. However, this survey will only cover low to
intermediate mass star-forming regions and it may not be possible to
determine a robust value of $\alpha$. For this, we may need to wait
for the exquisite resolution and sensitivity of ALMA to probe nascent
regions of massive star formation. 

\acknowledgments 
We would like to thank Brendan Bowler, John Johnson, and Norbert
Sch\"{o}rghofer for their input and contributions to this research. We
are also grateful for comments on our first draft provided by
E. Rosolowsky.

\end{document}